# TARGET SELECTION FOR SETI: 1. A CATALOG OF NEARBY HABITABLE STELLAR SYSTEMS

A Catalog of Nearby Habitable Stellar Systems

MARGARET C. TURNBULL
University of Arizona, Steward Observatory, 933 N. Cherry Ave., Tucson, AZ 85721, (520) 621-8125, turnbull@as.arizona.edu

AND

JILL C. TARTER
SETI Institute, 2035 Landings Dr., Mountain View, CA 94043, tarter@seti.org

ABSTRACT
In preparation for the advent of the Allen Telescope Array, the SETI Institute has the need to greatly expand its former list of ~2000 targets compiled for Project Phoenix, a search for extraterrestrial technological signals. In this paper we present a catalog of stellar systems that are potentially habitable to complex life forms (including intelligent life), which comprises the largest portion of the new SETI target list. The Catalog of Nearby Habitable Systems (HabCat) was created from the Hipparcos Catalogue by examining the information on distances, stellar variability, multiplicity, kinematics and spectral classification for the 118,218 stars contained therein. We also make use of information from several other catalogs containing data for Hipparcos stars on X-ray luminosity, CaII H&K activity, rotation, spectral types, kinematics, metallicity, and Strömgren photometry. Combined with theoretical studies on habitable zones, evolutionary tracks and third body orbital stability, these data were used to remove unsuitable stars from HabCat, leaving a residue of stars that, to the best of our current knowledge, are potentially habitable hosts for complex life. While this Catalog will no doubt need to be modified as we learn more about individual objects, the present analysis results in 17,129 Hipparcos "habstars" near the Sun (75% within 140 pc), ~2200 of which are known or suspected to be members of binary or triple star systems.

Subject Headings: Extraterrestrial intelligence --- astrobiology --- solar neighborhood

§1. INTRODUCTION

§1.1 Motivation for HabCat: Project Phoenix and the Allen Telescope Array

The creation of a Catalog of Habitable Stellar Systems (HabCat) was motivated specifically by a need for an expanded target list for use in the search for extraterrestrial intelligence by Project Phoenix of the SETI Institute. Project Phoenix is a privately funded continuation of NASA's High Resolution Microwave Survey (HRMS), a mission to search for continuous and pulsed radio signals generated by extrasolar technological civilizations. HRMS consisted of an All Sky Survey in the 1 to 10 GHz frequency range as well as a Targeted Search of 1000 nearby stars at higher spectral resolution and sensitivity in the 1 to 3 GHz range. Although Congress terminated HRMS in 1993, the SETI Institute raised private funds to continue the targeted portion of the search as Project Phoenix. Project Phoenix now carries out observations at the Arecibo Observatory in conjunction with simultaneous observations from the Lovell Telescope at the Jodrell Bank Observatory in England. The project uses a total of three weeks of telescope time per year and is able to observe ~200 stars per year.

In the near future, the SETI Institute expects to increase the speed of its search by a factor of 100 or more. In a joint effort by the SETI Institute and the University of California-Berkeley, the Allen Telescope Array (ATA, known formerly as the One Hectare Telescope) is currently being designed for the Hat Creek Observatory located in northern California. The ATA will consist of 350 dishes, each 6.1 meters in diameter, resulting in a collecting area exceeding that of a 100-m telescope. On its current development and construction timeline, the ATA should be partially operational in 2004 and fully operational in 2005. The construction of the ATA will mark an increase in telescope access and bandwidth capability sufficient to observe thousands to tens of thousands of SETI target stars per year. Hence the observing list for Project Phoenix needs to be greatly expanded from its original scope of about 2000 of the nearest and most Sun-like stars (Henry et al. 1995). The Catalog of Nearby Habitable Stellar Systems (HabCat) presented in this paper comprises the largest portion of SETI's new target list (to be discussed fully in a subsequent publication).

§1.2 Defining Habitability

Our goal is to build a catalog of stars that are potentially suitable hosts for communicating life forms. In defining the habitability criteria for SETI target selection, we note that the development of life on Earth required (at the very least) a terrestrial planet with surficial liquid water and certain heavy elements (e.g., phosphorus), plus an energy source (e.g., Sunlight) (Alberts et al. 1994). The basic requirement of terrestrial planets suggests that there may be a lower limit on stellar metallicity for habitability (discussed in §3.7). Given the possibility of terrestrial planets, the second requirement of liquid water means that the concept of a "habitable zone" (HZ), i.e., that annulus around a star where the temperature permits the presence of liquid water on an Earthlike planet (investigated in detail by Kasting, Whitmire & Reynolds 1993), is a recurring theme as we evaluate the habitability of stars with different spectral types and also multiple star systems.

An additional requirement for the development of *complex* life on Earth has been the continuous habitability of the planet over billions of years. Although there is evidence that simple life forms inhabited Earth as early as 0.8 billion years after Earth's formation (Schopf 1993), according to the fossil record and biomolecular clocks, multicellular life did not appear until after 3-4 billion years (e.g., Rasmussen et al. 2002; Ayala, Rzhetsky, & Ayala 1998; Wray, Levinton, & Shapiro 1996), and the emergence of a technological civilization capable of interstellar communication occurred only in the last century. The requirement for a long habitability timescale, $\tau_{hab}$, strongly impacts the number of stars included in HabCat. All such stars must be older than $\tau_{hab}$, and their HZ locations must not change by more than the HZ width over that time. However, it is not clear that the 4.6 billion year time to intelligence on Earth is a universal requirement for the appearance of interstellar communication technology (e.g., arguments made by McKay 1996). Here we acknowledge that the determination of a minimum $\tau_{hab}$ for SETI targets is arbitrary, and we follow the examples of Dole (1964), Hart (1979) and Henry et al. (1995) in setting $\tau_{hab}$=3 billion years.

Combining these ideas, we define a "habitable" stellar system as a system in which an Earthlike planet could have formed and supported liquid water throughout the last 3 billion years. For convenience, we call the host star of such a system a *habstar*. Implicit in the definition of a habstar are concerns about metallicity, companions, stellar age and mass, and stellar variability. We expect that the habitability criteria presented below will need to be adjusted as more is learned about the formation of terrestrial planets, the origin and evolution of life on Earth, and the presence of life, if any, on other planets or moons of the Solar System.

§1.3 The Hipparcos Catalogue as a starting point for HabCat

The Hipparcos and Tycho Catalogues (ESA 1997a) meet many of SETI's needs in terms of astronomical data compilation. The Hipparcos mission's typical parallax standard errors of ~1 milliarcsecond (mas) allow unprecedented accuracy in distance measurements (standard errors of 20% or less for ~50,000 stars) and hence vast improvements in luminosity determinations, which is perhaps the most important information we use in determining the habitability of nearby stars. The Catalogue also includes accurate photometry (B−V uncertainties typically less than 0.02 magnitudes; important for determining bolometric corrections and habitable zone locations) and proper motion data (which we use to constrain membership to the Galactic disk), as well as information on variability and multiplicity. All these data contribute to the goal of creating a sample of well-characterized stars that we believe may provide suitable habitats for complex life forms. However, the data used in this paper regarding stellar kinematics, chromospheric activity, rotation, metallicity, multiplicity, etc, are still missing for most stars in the Solar Neighborhood. Therefore our procedure was to begin with the entire Hipparcos Catalog and eliminate stars for which currently available data indicate *non*-habitability.

§1.4 The Creation of HabCat

The creation of HabCat was carried out in two phases. First, the Celestia 2000 program published on CD-ROM by the European Space Agency (ESA 1997b) was used to query the Hipparcos database and exclude undesirable stars using the flags present in the Hipparcos Catalogue itself. Celestia 2000 allows the user to specify many criteria and combine them to generate a subset of stars. The resulting subset can be output in machine-readable format with all of the information available from the Hipparcos Input Catalogue, the Hipparcos Catalogue and the Tycho Catalogue. We refer to this subset generated with our habitability criteria as the "Celestia sample." In the second phase of target selection, stars were removed from the Celestia sample using information from external databases or other calculations, so as to keep only stars which appear to be potential habstars based on spectral type, age, variability, metallicity or stellar/substellar companions.

In Section 2 of this paper we present the criteria used for the creation of the initial Celestia sample, and in Section 3 we describe additional cuts made by matching these Hipparcos stars with data from other databases. In Section 4 we present the resulting Catalog of Nearby Habitable Stellar Systems.

§2. THE CELESTIA SAMPLE

§2.1 Minimal Data Requirements and Uncertainty Limits

In order to do any assessment of the habitability of Hipparcos stars, we require some estimate of luminosity and temperature. Therefore we have only included stars for which B and V photometry and parallax were obtained, and we did not include stars whose parallax measurements were less than zero (due to large uncertainties). Additionally, the quoted fractional parallax uncertainties (the mean standard error in the mean) $\sigma_\pi/\pi$ were limited to 30%, which corresponds to a range of −0.77 mag to +0.57 mag around the calculated $M_v$. This uncertainty in visual luminosity is acceptable because we expect a solar mass star to increase in brightness by ~1 mag between the zero-age main sequence and the Hertzsprung gap on the HR-diagram, so we can still determine whether or not our stars are on the main sequence. The uncertainty in locating stars to the main sequence is dominated by that of the parallaxes measurements, and photometric uncertainties are generally small for the vast majority of stars in the Hipparcos Catalogue (median precision 0.06 mag in V, 0.07 mag in B).

§2.2 Variability Detected by Hipparcos

What about variable stars? We know that all stars are variable at some level, but how much fluctuation is tolerable to life, or to complex life? The most well-studied variation of the Sun is the 11 year Sunpot cycle (also called the Schwabe cycle), and during Solar Minimum the Sun's total irradiance fluctuates by only ~0.02%. During Solar Maximum, the Sun's output appears to increase both in total irradiance (by ~0.1%, corresponding to a global terrestrial tropospheric temperature increase of 0.5-1.0º C) and in stochastic variability (to ~0.15%) (Willson & Hudson 1991; Lean 1997 and references therein). The Solar wind, UV and X-ray fluxes, and coronal mass ejection events also increase in intensity during Maximum, and the resulting geomagnetic storms, drop in cosmic-ray fluxes, and change in ozone production impact terrestrial atmospheric circulation, temperatures and weather (Haigh 1996, Cliver 1995, Tinsley 1998). There appears to be a longer and more pronounced variation cycle superimposed on the Schwabe cycle (the ~200 year "Suess wiggles", Damon & Sonnett 1991), and the most recent minimum in this cycle occurred between 1645 and 1715 AD (the Maunder Minimum) with Solar irradiance levels dropping by 0.22–0.55% from the current quiet Sun (Baliunas & Jastrow 1990; Wigley & Kelly 1992). This decrease in Solar irradiance coincides with the coldest years of the Little Ice Age of 1550-1700 AD. Likewise, the Medieval Maximum of sunspot activity (1100 to 1250 AD) coincides with a warm interval on Earth known as the Medieval Warm Period. These Solar variations were not deleterious to complex biology, nor did they prevent the emergence of technological civilization. However, given that these extremely small fluctuations did have noticeable impacts on global climate, we have taken the view that stellar variability greater than ~1% in luminosity would be a significant concern for habitability.

The Hipparcos mission was able to detect flux variations of ~3% in the Hp bandpass during the four year mission, a level of variability at least 5 times greater than the Sun's change in output since the Maunder Minimum and possibly beyond the limit of variability tolerated by complex life forms. However, the fluctuations detected by Hipparcos are on a much shorter timescale than those of the Sun and thus may not be completely analogous (e.g., the thermal inertia of a planet's atmosphere and oceans may filter out short period fluctuations so that only the average luminosity is relevant to climate). Erring on the conservative side, we have chosen to remove all stars with detected variability: "unsolved" variables (flagged as "U" in column H52 or "R" in H52 and "2" in H53), "microvariables" (flagged as "M" in H52), "variability-induced movers" (flagged as "V" in field H59), and stars included in the Variability Annex Part C ("C" flag in H54, nonperiodic or unsolved variables) were omitted during the Celestia query. As for periodic variables (flagged as "P" in H52), stars classified as cataclysmic, eruptive, pulsating, rotating or X-ray variables were eliminated, but variables identified as eclipsing binaries were retained for analysis in §3.8.

§2.3 Multiplicity in the Hipparcos Catalogue

Approximately 20% of the Hipparcos stars are known or suspected to be members of a multiple star system. There are undoubtedly many more multiple systems in the Hipparcos Catalogue that have not been identified, and studies of the Solar Neighborhood suggest that the fraction of Solar-type stars in binary or multiple systems is closer to 2/3 (Abt & Levy 1976, Duquennoy & Mayor 1991). The presence of more than one star in a stellar system places limitations on where planets can form and persist in stable orbits. In order for a multiple system to be habitable to life, stable planetary orbits must coincide with the habitable zone. In the creation of HabCat, we have examined multiple systems individually for habitability by matching HZ location calculations with planetary stability zone locations (discussed in §3.8). Eclipsing binaries, spectroscopic binaries, astrometric binaries, and visual binary and multiple systems where no more than 2 components were resolved in each associated Hipparcos entry were retained for this analysis. We excluded from the Celestia sample entries containing more than 2 resolved components (indicated in field H58). Stars with "stochastic" astrometric solutions (flagged as "X" in H59) are likely to be astrometric binaries with periods less than ~3 years, and they were also excluded from the Celestia sample.

§2.4 The Celestia Query and Resulting Sample

The Celestia query resulted in a total of 64,120 stars out of the original 118,218. In Table 1, we show the exact criteria specified in the Celestia query. Figure 1 shows the resulting color-magnitude diagram. For a given star, the B- and V-band photometry listed in the Hipparcos Catalogue (and plotted in our Figures) represents either ground-based measurements (collected from the literature), the converted Hipparcos mission measurements (where the Hp magnitude was converted to Johnson-Cousins V magnitude) or the converted Tycho measurements (where $B_T$ and $V_T$ were converted to Johnson-Cousins B and V magnitudes). While the quoted photometric uncertainties are small (less than 0.02 mag in B−V for 75% of Celestia sample stars), Bessell (2000) has noted that the published converted Tycho measurements differ slightly from Johnson-Cousins measurements by up to 0.05 magnitudes, and that the residuals are a slowly varying function of B−V. Despite this, the uncertainty in position on the color-magnitude diagram is dominated by the uncertainty in distance for 98% of Celestia stars and the photometric errors do not affect the analysis carried out below.

From the color-magnitude diagram presented in Figure 1 it is obvious that there are many giants present in the Celestia sample, a few white dwarfs, and many stars of early spectral type (O-A) on the main sequence. In the analysis below we use our timescale for habitability, $\tau_{hab}$, to define a "locus of habitability" on the color-magnitude diagram. After removing stars outside this locus and applying other cuts we will look at the final distribution of spectral types in HabCat.

Another noticeable feature of the Celestia color-magnitude diagram is the enhanced number of objects located inside the region $1.55 < B-V < 1.75$ and $6.5 < M_v < 9.5$. The initial concern was that these are pre-main sequence stars, which would certainly violate our $\tau_{hab}$ criterion for minimum age. Rather, this scatter above the lower main sequence is caused primarily by objects that had no available photometry at the time of the creation of the Hipparcos Input Catalog and were below the detection threshold for Tycho photometry. These objects appear systematically too red and are likely to be main sequence K stars. We have drawn our "terminal-age main sequence" cut in §3.1 so as to include these stars in the sample.

Finally, we note that many stars in the Celestia sample are known or suspected members of a multiple system on the basis of one or more indicators: 9421 are associated with a Catalog of Components of Double and Multiple systems (CCDM) identifier, 1986 required an astrometric acceleration term, 201 have orbital astrometric solutions, 414 have an eclipsing binary-type light curve, and 1856 are associated with a "suspected non-single" astrometric solution quality flag. These non-exclusive categories affect a total of 12,958 Hipparcos entries. Although our query specified that the number of resolved stellar components within one entry was not to exceed 2, there are some systems with more than 2 components contained in the sample where separate entries are related via the same CCDM identifier. The analysis of all binary/multiple systems for habitability is described in §3.8.

§3. CRITERIA FOR HABITABILITY: NON-HIPPARCOS DATA

§3.1 Habitability on the color-magnitude diagram

With $\tau_{hab}$ = 3 billion years, HabCat is limited to low mass stars on the main sequence. As stars evolve off the zero-age main sequence toward luminosity class IV, the HZ moves slowly outward and may encompass planets that were previously too cold to be habitable (e.g. Lorentz, Lunine & McKay 1997). Meanwhile, planets that were once in the star's habitable zone will suffer runaway greenhouse effects and lose any liquid water present (see Kasting et al. 1993). One can imagine that an advanced civilization could relocate to a more temperate world and persist throughout a star's main sequence lifetime. Therefore we consider a habitable system to remain habitable from 3 Gyr until it reaches the "terminal-age" main sequence, even though the HZ may move more than its own width during that total time (as is true for the Sun). We define the "terminal age" of the main sequence (TAMS) to be at the Hertzsprung gap in the color-magnitude diagram, where the lack of stars is due to large changes in effective temperature (and stellar radius) over timescales much shorter than $\tau_{hab}$. In order to determine what range of masses is acceptable for HabCat, we have used the TYCHO stellar evolution program (described by Young et al., 2001) to evolve Solar metallicity models of stellar masses between 1 and 2 $M_\odot$. We computed stellar age starting from the onset of core hydrogen burning, and a star that reaches the TAMS before 3 billion years is not considered habitable. With TYCHO models we find that a star of ~1.5 $M_\odot$ (spectral type ~F5) is just habitable and that stars of this mass reach a maximum absolute brightness of $M_v$~2.5 before approaching the TAMS.

Using this information we have applied two cuts to the Celestia sample: (1) a cut based on color-magnitude data and (2) a cut based on spectral types listed in the literature. The first cut removed all stars that are located above the TAMS drawn in Figure 1 (long-dashed curve, described by $M_v \leq -10 ((B-V) - 1.4)^2 + 6.5$, plus a B−V < 1.75 requirement), all stars brighter than $M_v$=2.5, or below the main sequence (short-dashed line, $M_v \geq 28 (B-V) + 5.8$ for B−V < −0.1; $M_v \geq 4.8 (B-V) + 3.5$ for −0.1 < B−V < 1.28; $M_v \geq 17 (B-V) - 12.2$ for B−V > 1.28). To apply a cut on spectral types, we matched each Celestia star to its corresponding two-dimensional MK spectral type (where available) in the catalogs of Houk and collaborators (Houk & Cowley 1975; Houk 1978; Houk 1982; Houk & Smith-Moore 1988; Houk & Swift 1999).  When a spectral type was not available from these catalogs, the spectral type listed in the Hipparcos Input Catalog was used.  We then removed stars with undesirable spectral types, including spectral types earlier than F5, stars with noticeable emission lines, nebulous lines, weak lines, variable lines and shell stars.  Stars with spectral types containing the following characters were removed: O, B, A, D, C, F0, F1, F2, F3, F4, II, III, Ia, Ib, VI, delta, sd, sh, W, e, var, R, N, S, w, v, or n.  The ~2% of Celestia stars that did not have spectral types listed in our sources were examined only according to color-magnitude location.  These two cuts resulted in the removal of 64% of the stars from the Celestia sample (leaving 23,246).

§3.2 The Habitability of F Stars

There are ~8600 F stars in the remaining list.  In addition to concerns about main sequence lifetime, the distribution of a star's spectral energy output is relevant to the evolution of Earth-like biology.  Biological molecules, including nucleic acids (found in DNA and RNA), amino acids (proteins) and lipids (cell membranes), absorb strongly at ultraviolet wavelengths (see Cockell 1999 and references therein), and each absorption event has the potential to disrupt the molecule.  Even with the sophisticated repair mechanisms commonly employed at the cellular level, the Sun's UV output would be fatal to complex life were it not for the Earth's protective ozone layer.  For F5V stars with effective temperatures near 6800 K and luminosities of ~4 $L_\odot$, UVC (200–280 nm) emissions can be ~10 times that of the Sun, depending on the extent of metal-line blanketing (see IUE spectra in Heck et al. 1984).  Given the biological sensitivity to these wavelengths, this raises the question of whether additional limits on spectral type should be imposed.

Kasting, Whittet and Sheldon (1996) addressed this issue and showed that, since ozone formation is initiated by the splitting of $O_2$ by ultraviolet photons, Earth-like planets orbiting F-stars may have thicker ozone layers and receive *even less* UV radiation at their surfaces than does the Earth.  For K stars the amount of UV radiation and hence ozone production is less, but the lack of incident radiation more than compensates for the lack of ozone shielding, and in fact it may be G-type stars that are the least habitable to complex life forms in terms of ultraviolet radiation.  Yet even G stars are capable of supporting technological civilizations. Therefore we have not culled our Habitable Systems Catalog on the basis of stellar UV emissions.

§3.3 The Habitability of M Stars

There are also ~600 M stars on our remaining list. Although this is a small fraction of the total list (~2.5%), the habitability of M dwarfs is of general interest to SETI because these stars may account for more than 70% of the stars in the Galaxy (Henry et al., 1999). Previous SETI target lists did not include M dwarfs for several reasons. The first concern was that the fully convective interiors and rapid rotation of some M stars give rise to highly energetic flaring events and that the ultraviolet emissions from these flares could be harmful to lifeforms in the nearby HZ. However, Doyle et al. (1991) assert that M star flares would generate large amounts of ozone in the upper atmosphere of a planet, which would in turn protect the planetary surface from the intensified UV radiation. Furthermore, they point out that the total amount of UV radiation from flares even at this small distance from the star is still less than the Solar UV radiation flux on Earth. Therefore we have not explicitly excluded M stars on the basis that some of them are flare stars. However, we note that we tried to avoid frequently flaring stars by excluding non-periodic variables in the Celestia query, and by excluding in §3.1 spectral classifications flagged with "e" (indicating the presence of Balmer emission lines which is generally associated with chromospheric activity, e.g. Mauas et al. 1997) in Houk's catalogs.

A second consideration for fainter stars is that of habitable zone width. Because of the inverse-square fall-off of radiation, the distance at which a planet in thermal equilibrium with its parent star will be at a given temperature goes as the square root of the stellar luminosity. Thus the total width of the habitable zone also goes as $L^{0.5}$ and the HZ narrows as we consider later spectral types. Kasting, Whitmire & Reynolds (1993, hereafter K93) estimate the HZ width (from "runaway greenhouse" to "maximum greenhouse") for the Sun to be ~0.83 AU, but for an M0 star the instantaneous HZ width is only ~1/3 of this and would seem to be 1/3 as likely to host a habitable planet. On the other hand, their long main sequence lifetimes suggest that a planet orbiting within an M dwarf HZ is more likely to eventually develop complex (and potentially communicating) life. To make a meaningful comparison of stars of different spectral types we need to factor in both HZ time and area by using K93's concept of the continuously habitable zone (CHZ), defined as the annulus that is continuously habitable over a specified amount of time. For $\tau_{hab}$~3 billion years, G stars have the largest CHZs (~0.5 AU), M0 stars have CHZs of just under half this width, and CHZs disappear altogether for spectral types earlier than our F5 cutoff. In these terms, M stars appear to be reasonable locations for the appearance of life.

The location of M star HZs may be a concern for habitability in that Earth-mass planets may not usually form at this small distance from the star (Wetherill 1996). Furthermore, K93 note that the CHZs of M stars are located within the tidal lock radius (and would therefore be synchronous rotators). They suggested that planets orbiting within the CHZ would have one side of the planet perpetually dark and cold enough to freeze out any atmosphere. Haberle et al. (1996) and Joshi et al. (1997) explored this issue and calculated that a $CO_2$ atmosphere of only 0.1 bars would be enough to circulate heat from the illuminated side of the planet and prevent atmospheric collapse, while 1.5 bars of $CO_2$ would allow temperatures compatible with the existence of liquid water over much of the surface of the planet.

Many questions remain regarding the habitability of M dwarf systems. The spectral energy distribution of these stars may have negative consequences for photosynthesis, and the thicker atmospheres required to prevent atmospheric collapse in synchronously rotating planets may interfere with the transmission of photosynthetically useful radiation over much of the planet's surface. Thus these stars may indeed be less attractive targets, but the potential problems are not necessarily insurmountable. Given the small number of M stars found in the Hipparcos Catalogue (and hence the small amount of telescope time devoted to them), we opt at this time to include them in HabCat, except where they fail the habstar criteria applied to all spectral types.

§3.4 Stellar Age Indicators: Chromospheric Emission

As part of a minimum age cut for HabCat, we have already removed from our target list those stars of spectral types known to have lifetimes shorter than the required 3 Gyr. But for stars of later spectral types, habitability determination on the basis of age is less straightforward. Henry et al (1995) proposed using the Ca II H and K line chromospheric activity indicator for age determination of potential SETI targets (as described in Noyes, Weiss & Vaughan 1984). Due to periodic activity cycles (like the 11 year Solar cycle), typical 1-$\sigma$ uncertainties for stars with B–V > 0.6 are about 0.4 to 0.5 dex in log(age) (Lachaume et al. 1999). This translates into an uncertainty of +/–3 Gyr for a 5 billion year old star. Over the past 400 years, the chromospheric activity of the Sun has ranged from log $R'_{HK}$ ~ −5.1 during the Maunder Minimum (corresponding to an estimated age of ~8 Gyr) to log $R'_{HK}$ ~ −4.75 during Solar Maximum (corresponding to an estimated age of ~2 Gyr) (Baliunas et al. 1995a, Baliunas et al. 1995b). Nevertheless, it seems that stars which show activity cycles always remain on either the "active" or "inactive" side of log $R'_{HK}$ ~ −4.75 (Henry et al. 1996). Thus it should be possible to tell from only one observation on which side of this line a star will continue to reside. According to the relation from Donahue (1993),

$$\log \tau = 10.725 - 1.334\, R_5 + 0.4085\, R_5^2 - 0.0522\, R_5^3,$$

where $\tau$ is the age in years and $R_5$ is defined as $R'_{HK} \times 10^5$. $R'_{HK} = -4.75$ corresponds to an age of ~2.2 Gyr. Thus for our purposes, we can rule out all stars with activity index log $R'_{HK}$ > −4.75. Figure 2 shows the distribution in $R'_{HK}$ for the 1408 stars in the sample resulting from §3.1 in comparison with Donahue's age-activity relation ($R'_{HK}$ data from T. J. Henry and D. R. Solderblom 2002, private communication). With these data 425 more objects were identified as "young" and cut, leaving 22,821 stars.

§3.5 Stellar Age Indicators: X-ray Luminosity

Another indicator of age is X-ray luminosity. Pre-main-sequence stars tend to emit X-rays in excess of that expected from the blackbody temperature deduced from their optical spectra, and these emissions steadily decrease with age. Güdel et al. (1997) investigated X-ray emissions for nine Sun-like stars of ages 70 Myr to 9 Gyr and found that

$$L_x \sim 2.1 \times 10^{28}\, \tau^{-1.5}\, [\text{ergs s}^{-1}],$$

where $L_x$ is the total luminosity in the 0.1–2.4 keV energy range (i.e., the energy range observed by the *ROSAT* Position-Sensitive Proportional Counter (PSPC) instrument) and $\tau$ is the stellar age in Gyr. In the *ROSAT* All Sky Survey (RASS), young active late-type stars (F through M) were found to account for 85% of the soft X-ray sources detected in the Galactic plane (Motch et al. 1997). Guillout et al (1999, hereafter G99) performed a cross-correlation of the RASS with the Hipparcos and Tycho Catalogues and found 6200 matches with Hipparcos entries. The X-ray/Optical offset was required to be less than 30 arcseconds, a separation for which the authors calculate a ~7% spurious match rate. In Figure 3 of G99 the PSPC count rate as a function of exposure time is shown for all of the RASS-Tycho matches. From this Figure we can see that most of the X-ray sources were detected at an exposure time of $t_{exp} \sim 450$ s, which corresponds to a detection threshold of $S_{thr} \sim 0.015$ counts/s. Combining this with the result above from Güdel et al. (1997), the detectable X-ray luminosity as a function of distance and age is then

$$L_x = 4\pi d^2 \, S_{thr} \times ECF = 2.1 \times 10^{28} \, \tau^{-1.5} [\text{ergs s}^{-1}] \,,$$

where ECF is the counts-to-energy conversion factor. The ECF is a function of spectral hardness and interstellar absorption, but G99 found a value of $10^{-11}$ erg cm$^{-2}$ s$^{-1}$ per count s$^{-1}$ to be a good average conversion factor for the RASS-Tycho and RASS-Hipparcos matches. Using this value, we calculate the maximum distance at which a star of a given age should be detectable to *ROSAT*:

$$d_{max}(pc) = 34 \times \tau \, (\text{Gyr})^{-0.75} \,.$$

This function is plotted in Figure 3 and the 3 Gyr detection limit of 14.8 pc is marked. Superimposed on Figure 3 is the normalized histogram for the 2127 RASS sources matched to the stars in our remaining sample. From this plot we can see that the detection limits of *ROSAT* PSPC allow us to probe distances up to ~200 pc for late spectral-type stars younger than 100 Myr. The vast majority of these detections are beyond ~15 pc and hence, according to the above calculations, younger than 3 Gyr, and we have removed all of them from our sample. The 120 stars that were detected at smaller distances are not necessarily this young, and they have been retained. Thus 20,814 stars remain in HabCat after cutting 2007 RASS sources. We note that the above age-$L_x$ relation was derived for G dwarfs, and lower mass stars may in fact take longer to decrease in X-ray luminosity (due to a longer spin-down time, see Soderblom et al. 1993) so that in terms of $L_x/L_{bolometric}$ they will appear younger than their Solar-mass counterparts of the same age. However, lower mass stars also put out less total luminosity, and in terms of detectability this offsets their larger fractional X-ray fluxes (see Figure 7 of G97).

§3.6 Stellar Age Indicators: Rotation

Ultimately, the X-ray emissions and chromospheric activity are associated with stellar rotation. Stars are born with relatively high rotation velocities (~100 km/sec) but lose angular momentum through magnetically driven winds (Soderblom et al. 1993). The rotation slows with time, leading to decreased X-ray emissions and chromospheric activity. Since the magnetic dynamo depends on the depth of the convective envelope, the relationship between rotation period and age also depends on stellar mass. Lachaume et al. (1999) combined the period-age-mass relation found by Kawaler (1989) and Barry (1988) with evolutionary tables from Bertelli et al. (1994) to obtain an estimate of stellar age as a function of rotation period, B−V color, and [Fe/H]. However, Tables A1 and A2 of Lachaume et al. show that this formula consistently gives ages that are lower than ages derived from Bertelli's isochrones by 2.5 Gyr or more. In light of this discrepancy we used rotation velocities only to exclude those stars that are unambiguously young. In order to avoid throwing out old stars (like υ And, $\tau_{isochrone}$~3 Gyr from Lachaume et al. 1999, $v \sin i$ ~ 9 km s$^{-1}$ from Glebocki et al. 2000), we have chosen to make a single cut in rotation velocity at $v \sin i$ = 10 km s$^{-1}$, consistent with ages much less than 1 Gyr (at least for G-type stars and later spectral types, Soderblom et al. 1993). Data were taken from the Glebocki et al. (2000) catalog of projected rotational velocities for 4865 Hipparcos stars, of which 426 stars remain in our list. Of these, 49 rotationally young stars were identified and eliminated, leaving a sample of 20,765 potential habstars.

§3.7.1 Habitability and metallicity

Although the presence of terrestrial planets has yet to be demonstrated for any main sequence star other than the Sun, stellar metallicity may indicate whether terrestrial planets are possible. Certainly there must be a lower limit in metallicity below which there are not enough heavy elements to build an Earth-mass planet during the system's formation. Recent work by Reid (2002b) shows that the frequency of detected planetary systems is clearly correlated with stellar metallicity. The Sun itself is typical in metallicity amongst G-type disk dwarfs within 25 pc (Reid, 2002b) but it is less metal-rich than ~2/3 of the stars with close-orbiting giant planets (Lineweaver 2001, Santos et al. 2001, Reid 2002a). If it is the case that metallicity is an indicator not only of planet occurrence but also of planet mass then there may be a limited range of metallicities where low mass (terrestrial) planets exist (Lineweaver 2001), which would necessitate both minimum and maximum metallicity cuts. As extrasolar planet detections accumulate we will be able to explore more thoroughly any relationship between metallicity and planet mass (and other parameters such as semimajor axis or eccentricity), but at this time there is no clear case for a correlation between planet mass and stellar metallicity (see data and analysis in Reid 2002a). Consequently, we conclude that a minimum metallicity cut is justified, but we chose not to impose an upper limit in metallicity.

The exact value in [Fe/H] that is theoretically too low for planet formation is difficult to estimate because of the many variables involved and uncertainties remaining in star and planet formation theory (e.g., whether a more massive circumstellar disk could compensate for lower metallicity, how disk parameters are correlated with stellar mass, and to what extent giant planet mass and position can alter terrestrial planet mass, number, and position (see Wetherill 1996)). We can get a rough estimate of minimum metallicity by assuming (1) that the total terrestrial planet mass scales linearly with the abundance of the initial star-forming cloud (which is reflected in the abundance of the observable central star), (2) that the mass of the terrestrial planets in the Solar System are is typical for Solar metallicity, and (3) that one Earth mass is the minimum amount of material necessary to create a habitable planet. Then the Solar System's total terrestrial planet mass (~2 Earth masses) implies a minimum metallicity for habitability of [Fe/H] ~ -0.3. In the next Section we describe an observationally convenient metallicity cut that is consistent with this estimate.

§3.7.2 Kinematics as an indicator of low metallicity

One tactic for identifying low-metallicity stars is to make use of kinematics data and the kinematics-age-metallicity relationship. Older generations of stars tend to be lower in metallicity and therefore they are expected to be less likely to harbor terrestrial planets. For stars in the galactic disk, a connection between metallicity and kinematics also exists because older stars are also more likely to have undergone gravitational encounters with other stars at some point during their lifetimes, and thus they tend to exhibit larger velocities relative to the local standard of rest (LSR). Stars that formed as part of the halo have even higher velocities and are thought to be the oldest existing population in the Galaxy, with near zero metallicity (see

Binney & Merrifield 1998, Chapter 10 and references therein).

To find general trends in metallicity vs. kinematics, we use the data presented by Edvardsson et al. (1993, hereafter E93) for 189 nearby F and G disk dwarfs. Figure 4 shows the U, V and W Galactic velocities as a function of [Fe/H] for the E93 stars (U increases away from Galactic center, V increases along Galactic rotation, and W increases towards the North Galactic Pole). The marked increase in V and W velocity dispersion at [Fe/H] ~ −0.4 makes this a convenient choice for a kinematical cut-off, and from our calculations above we believe this metallicity is an acceptable lower limit for Solar System-like planet formation. (This metallicity is also chosen by Binney & Merrifield (1998) as the division between the "thin" and "thick" disk populations.) From E93, we find that for the 131 stars with [Fe/H] > −.4, the average velocities and dispersions in Galactic coordinates are:

$$\langle U \rangle = 1.0 +/- 3.0 \text{ km/sec}, \quad \sigma_u = 34.3 \text{ km/sec},$$
$$\langle V \rangle = -12.5 +/- 2.0 \text{ km/sec}, \sigma_v = 21.5 \text{ km/sec, and}$$
$$\langle W \rangle = -1.7 +/- 1.6 \text{ km/sec}, \sigma_w = 17.9 \text{ km/sec},$$

where the velocities have been transformed to the local standard of rest (LSR) by accounting for the Solar velocity, (U=−10.0 +/− 0.4, V=5.3 +/− 0.6, W=7.2 +/− 0.4) km/sec (Denhen & Binney 1998). We have used these mean values plus and minus three times the velocity dispersions as maximum and minimum velocity limits for potential habstars. The resulting ellipsoid of space velocities included in HabCat is then described by:

$$(U - 1 \text{ km/sec})^2/(3 \times 34.3 \text{ km/sec})^2 +$$
$$(V + 12.5 \text{ km/sec})^2/(3 \times 21.5 \text{ km/sec})^2 +$$
$$(W + 1.7 \text{ km/sec})^2/(3 \times 17.9 \text{ km/sec})^2 \leq 1.$$

To calculate U, V and W space velocities for our list of potential habstars, we have used coordinates, proper motions and parallaxes from the Hipparcos Catalog and radial velocities from Barbier-Brossat & Figon (2000, hereafter BBF). Only data where $|v_{rad}| > 3\ \sigma_{vrad}$ were used, and 1339 stars from the BBF catalog were thus matched to our remaining list of 20,765 stars. For stars without radial velocity data, $v_{rad}$ was assumed to be zero, and the U, V and W velocities calculated are lower limits. We did not attempt to correct for Galactic rotation because all of our stars are within ~300 pc, and recent determinations of the Oort constants (e.g., Feast & Whitelock 1997, A = 14.8 +/− 0.8 km/sec/kpc, B = −12.4 +/− 0.6 km/sec/kpc) imply that the total variation in the velocity of the local standard of rest over this space is only on the order of a few km/sec, a small fraction of our velocity ellipsoid.

Thus we have separated our stars into two populations: "kinematically high metallicity" (hereafter KHM) stars and "kinematically low metallicity" (hereafter KLM) stars. The velocity space occupied by KHM stars (black points) and KLM stars (red points) is shown in Figures 5a and 5b (as above, the velocities are corrected for Solar motion), and using this criterion we have removed 1917 additional stars from HabCat (leaving 18,848).

We can test whether our kinematic cut is in fact removing low metallicity stars by looking at the cut for stars that have both radial velocity data and metallicity data. Figure 6 shows, for 699 KHM stars and for 209 KLM stars (all of which have radial velocity measurements), the distribution of metallicities derived from either spectroscopic or photometric measurements (described in the next §3.7.3). We find that, as expected, there is an offset in mean metallicity for the two groups (<[Fe/H]> = −0.13 for the "high metallicity" sample and <[Fe/H]> = −0.64 for the "low metallicity" sample), and that the KLM sample includes many objects at extremely low metallicities.

However, we also find that the KHM group includes 68 stars with measured metallicities below our desired cutoff at [Fe/H] = −0.4 (vertical dashed line), and there are 83 stars in the KLM group with [Fe/H] ≥ −0.4. Therefore we expect that less than 10% of the stars predicted to be high metallicity will in fact have [Fe/H] < −0.4. We have also chosen to remove those 68 KHM stars that we know fail the metallicity cut in the next Section. For the 83 high metallicity stars with large Galactic velocity components, we point out that there may be a second reason to exclude from HabCat stars which have high velocities relative to the local standard of rest (LSR), regardless of metallicity: the LSR is believed to be nearly in co-rotation with the Galaxy's spiral pattern (Lépine et al. 2001, Balázs 2000). The co-rotation zone in the Milky Way has been proposed to give rise to a "Galactic Belt of Life," where stars near the Sun's Galactocentric radius with low velocities relative to the LSR will encounter spiral arms perhaps only once in several billion years (Balázs 2000, Marochnik & Mukhin 1988, Doyle & McKay 1991). The velocities of KLM stars depart significantly from that of the LSR and these stars are likely to be on elliptical orbits that bring them into more frequent spiral arm-crossings, where they encounter biologically damaging radiation fluxes perhaps $10^7$ times that currently received by the Earth (Clark et al. 1977, Doyle & McKay 1991). This criterion deserves further consideration as more is learned about the true structure of the Galaxy. For now, we have chosen not to include KLM stars, even where they are known to have [Fe/H] > −0.4.

§3.7.3 Spectroscopic and Photometric Measurements of Metallicity

We have also used direct spectroscopic measurements of stellar metallicity (Cayrel de Strobel et al. 2001, with data for 352 stars in our remaining sample) and metallicity estimates from Strömgren uvby photometry (Olsen 1983, Olsen & Perry 1984, Olsen 1993, Olsen 1994a, Olsen 1994b, with data for 6272 additional stars in our remaining sample) to identify and cut low metallicity stars. In keeping with the above Sections we have set [Fe/H] = -0.4 as the lower metallicity limit for HabCat. We used Schuster & Nissen's (1989) calibrations for F stars (0.22 < (b−y) < 0.375) and G stars (0.375 < (b−y) < 0.59) to estimate metallicities from Strömgren photometry. Stars with b−y colors outside this range were not considered. To check that the photometric metallicity estimates ([Fe/H]$_{phot}$) were close to the spectroscopically determined metallicities ([Fe/H]$_{spec}$), we compared the [Fe/H] estimates for 602 Hipparcos stars included in both the Cayrel de Strobel et al. and Olsen et al. catalogs. Figure 7 shows that while the photometrically determined metallicities depart from the spectroscopic measurements at very low metallicities ([Fe/H]$_{spec}$ < −2), [Fe/H]$_{phot}$ is generally good enough to identify stars with [Fe/H] < −0.4. Use of these catalogs has eliminated an additional 841 low metallicity stars from HabCat.

To test whether our metallicity cut is serving the purpose of eliminating stars which are not likely to host planetary companions, we have compared our lists of high and low metallicity stars to the list of stars known to host extrasolar planets. At the time of this writing, 80 Hipparcos stars are confirmed candidates for hosting extrasolar planetary systems based on radial velocity variations (http://www.exoplanets.org). Of these, 4 stars have metallicities below our [Fe/H] cutoff (HIP 5054, 26381, 64426, 64459). One explanation for low metallicity, giant exoplanets may be that they were formed by gravitational instabilities, rather than by core accretion and subsequent addition of a dense atmosphere (Boss 2002). In that case, these low metallicity systems are unlikely to harbor terrestrial planets or giant moons suitable for complex life. Furthermore, the analysis of planet-bearing stars by Reid (2002b) shows that the frequency of currently detectable planets drops from ~50% at [Fe/H]=0 to ~2% at [Fe/H]=−0.4. This means that adjusting our cuts to include the lowest metallicity of planet-candidate stars (HIP 64426, [Fe/H]~−0.7 from Cayrel de Strobel et al.'s 2001 catalog) would add fewer than 50 planetary systems to HabCat while increasing the number of stars in the catalog by ~2500. At this time we opt not to alter our bulk metallicity cut, even for cases in which low metallicity stars are known to harbor giant planets. This decision will be revisited if future observations provide evidence of terrestrial planets in low metallicity systems.

As a result of applying the low metallicity cut at [Fe/H] = −0.4, 18,007 potential habstars remain. The 55 remaining exoplanetary systems are assessed for dynamical habitability in §3.8.4.

§3.8.1 Habitability and Stellar Multiplicity

In terms of habitability, binary and multiple star systems are especially interesting because there is the potential for a habitable zone surrounding each star in the system as well as a circumbinary habitable zone. It is clear that, at least in certain configurations, planets can form and persist for billions of years in binary systems without being ejected or accreted (e.g., planets orbiting in double systems 16 Cyg B, 55 Cnc, $\tau$ Boo, $\upsilon$ And, and triple system HD 178911). However, not all binary orbits will permit dynamically stable habitable zones. Even if a dynamically stable HZ does exist, a planet orbiting there may receive wildly varying levels of radiation due to the constantly changing distance of the second star. Given that ~2/3 of stars are in binary or multiple systems (Abt & Levy 1976, Duquennoy & Mayor 1991), these issues are important to the overall habitability of the Galaxy. In the following Section we describe the analysis of habitability for Hipparcos binary and multiple systems in detail.

After all of the cuts described in the previous Sections, there remain 3507 unique entries in the HabCat that are known or suspected members of multiple star systems. This includes 2433 entries that are associated with an identifier from the Catalog of Components of Double and Multiple Stars (the "CCDM" category), 19 known eclipsing binaries ("EB"), 49 known spectroscopic binaries ("SB," data provided by D. Pourbaix, private communication), 627 entries flagged as suspected non-single stars ("SNS"), 37 orbital astrometric binaries with full or partial orbital solutions listed in the Hipparcos Double and Multiple Systems Annex ("O"), and 542 entries whose astrometric solutions required "acceleration" terms ("G"). Inventories for these non-exclusive multiplicity categories are listed in Table 2.

Before the dynamical and radiation variability analyses were performed, the list of binary/multiple systems was prepared as follows:

1. Hipparcos parallax and proper motion measurements were compared for the visual doubles/multiples where both components remain on our list of potential habstars. In this manner we identified CCDM visual doubles that are not physical binaries as evidenced by Hipparcos parallax measurements. However, we note that the analysis of parallax and proper motion data was not always carried out when Hipparcos observed only one component, or when only one CCDM component remained on our list of potential habstars. Therefore the cuts of this Section have likely excluded from HabCat a few single stars on erroneous grounds.

2. Wherever a CCDM identifier was associated with more than one Hipparcos entry, all components were removed if one or more components failed any of the criteria applied in the above Sections.

3. Stars whose astrometric solution quality was flagged as "suspected non-single" were removed, because we require an estimate of semimajor axis or period in order to assess habitability. Stars whose astrometric solutions required acceleration terms were not removed, as they are probably wide binaries (with periods much longer than the Hipparcos mission duration) whose orbits are unlikely to interfere with the habitable zones lying within a few AU of either star. The unseen component is assumed to be a main sequence star that is fainter than Hipparcos' detection limits. Eclipsing binaries, of which the longest period is 6.7 days, also do not affect their surrounding habitable zone, so none of these systems were cut.

4. Stars without angular separation measurements listed in either the CCDM or Hipparcos catalogs were removed from the list.

5. Each entry was examined by hand to identify visual triple systems. The CCDM catalog often lists only two components, while Hipparcos in fact resolved one of those CCDM components into two stars, resulting in a known triple system. There were no such known quadruple stars.

6. For relative simplicity, systems having more than two components were eliminated, except in the cases where (a) one component of a wide binary was resolved into two stars by Hipparcos, (b) one component of a visual double is an eclipsing binary, or (c) the visual double has a long period component as evidenced by astrometric acceleration terms. In case (a), all three components were taken into account when assessing the dynamical stability of the HZs as described in §3.8.2.

7. Each system was matched with V-band magnitudes and angular separations from the CCDM catalog. When the CCDM and Tycho photometry did not agree, Tycho photometry was used. Where Tycho photometry was available for only one component of a binary, the CCDM magnitude differences were used to calculate the magnitude of the second component. Where Tycho photometry represents the combined light of two components, the CCDM magnitudes were used to deconvolve the Tycho photometry into two magnitudes. For triples, V magnitudes for the third components (usually not observed by Hipparcos) were likewise calculated by using the CCDM magnitude differences. Where one component was resolved into two stars the separation was taken from the Hipparcos Catalogue, but V-band photometry for the individual components was generally not available in any catalog. In this case, individual V magnitudes were estimated from the Hp magnitude difference.

The above preparations removed 410 Hipparcos entries. For the remaining stars, habitable zones of visual double/triples were assessed for both dynamical stability and irradiance variations (§3.8.2), while spectroscopic and astrometric binaries were assessed only for dynamical stability because the magnitude of the secondary star is not known (§3.8.3). When one entry was associated with more than one multiplicity category, the star was assessed for habitability independently in each category.

§3.8.2 Visual Doubles/Multiples

Stable planetary orbits for a binary star can take two forms: S-type orbits, in which the planet orbits as a "satellite" to one of the stars, or P-type ("planetary") orbits, in which the planet orbits the whole binary. The question of habitability for these systems depends on whether the habitable zone set up by either an individual star or the binary as a whole resides within a region of dynamical stability. Figures 8a and 8b show the possible configurations of dynamically stable habitable zones of double and triple star systems, respectively. The dashed lines denote the "critical" semimajor axes for planets orbiting the primary and secondary, *inside* of which a planet will be dynamically stable, while the solid line denotes the critical semimajor axis for a planet orbiting the entire binary, *outside* of which a planet will be dynamically stable. The location of each critical distance depends on binary mass ratio, eccentricity and separation. For a binary to be habitable, one of the three possible HZs must be dynamically stable. We note that while the Figures portray several dynamically stable habitable zones, it is never true that circumstellar and circumbinary habitable zones simultaneously exist (i.e., as we move two stars closer together, their HZs eventually merge and become one circumbinary HZ).

We begin our analysis by locating the habitable zone for each system. For the outer edge of the HZ we have chosen to use the "maximum greenhouse limit" from Kasting, Whitmire, & Reynolds (1993, K93), which is the maximum distance from the star where a cloud-free $CO_2$ atmosphere can maintain a surface temperature of 273 K. Beyond this distance, additional $CO_2$ does not further warm the planet because the atmosphere is already entirely opaque in the infrared and the additional $CO_2$ serves to raise the planetary albedo through increased Rayleigh scattering. However, the maximum greenhouse limit may be conservative, because $CO_2$ clouds are expected to condense interior to this radius, and (unlike water clouds) their net effect would be to warm the planetary surface through a scattering greenhouse effect, thereby increasing the distance of the outer HZ boundary (Forget & Pierrehumbert 1997). In this analysis, the inner habitable zone limit was not needed, because even the closest visually resolved binaries had separations of more than a few AU, so circumbinary habitable zones were never possible.

Using these criteria, the outer limit of the Sun's HZ is ~1.67 AU with an atmospheric $CO_2$ partial pressure of ~10 bars. For stars of different spectral types, the radiation field will differ in total energy flux and in wavelength distribution, both of which have effects on the habitable zone limits. For a main sequence star more massive than the Sun the increase in stellar luminosity pushes the habitable zone outward, but the shifting of the spectrum towards the blue lessens this effect somewhat because the $1/\lambda^4$ dependence of Rayleigh scattering increases the effective planetary albedo. For a main sequence star less massive than the Sun, the smaller luminosity causes the HZ to move inward, but the shifting of radiation towards the red lessens this effect in two ways: (1) Rayleigh scattering is decreased thereby lowering the effective planetary albedo, and (2) the amount of starlight absorbed by the planet is further enhanced by high opacities in the near-infrared due to atmospheric $H_2O$ and $CO_2$. K93 included these effects in calculating HZ boundaries for stars of spectral types M0, G2 and F0. We corrected K93's HZ locations based on updated luminosity estimates for their listed effective temperatures (main sequence luminosities taken from Cox 2000). We then interpolated by eye for spectral types between F0 to M0 and extrapolated by eye down to our faintest stars at $M_v \sim 18$. We fit polynomials to the resulting data to obtain the approximate location of the habitable zone outer limit in terms of $M_v$:

$M_v < 3.46$:
$a_{out} = 23.21 - 19.41 \times M_v + 12.49 \times M_v^2 - 5.165 \times M_v^3 + 1.051 \times M_v^4 - 0.07996 \times M_v^5$

$M_v \geq 3.46$:
$a_{out} = 6.959 - 1.896 \times M_v + 0.2283 \times M_v^2 - 0.01433 \times M_v^3 + 0.0004473 \times M_v^4 - 5.393 \times 10^{-6} \times M_v^5$

The next step in identifying habitable binaries is to calculate for each system where stable planetary orbits are located. Holman & Wiegert (1999, hereafter H99) have investigated planetary stability in binaries with numerical simulations for a range of binary mass ratios and eccentricities. Our estimates of critical semimajor axes, $a_{crit}$, are very rough, as we do not have any direct measurement of mass ratios, orbital eccentricity, inclination or semimajor axis for visual binaries/triples. We estimated stellar mass ratios from V-band photometry, and for eccentricity we used $e = 0.67$, the average value for binaries with periods greater than 1000 days (Duquennoy & Mayor 1991). We also took the observed linear separation to be the semimajor axis. The distance $a_{out}$ was then required to be less than $a_{crit}$ for at least one star in order for a double system to be habitable.

As is shown in Figure 8b, the situation is slightly more complicated for triple systems. There is now an additional circumbinary radius *outside* of which planets will not be stable due to the presence of the third star, and this may eliminate the possibility of circumbinary stable orbits and interfere with S-type orbits inside the close pair. With the observed separations and our assumption of e = 0.67, there was no situation where a circumbinary or circumtriple habitable zone was possible.

In doing this HZ stability analysis we eliminated 310 Hipparcos entries, leaving 1387 entries that are associated with visual binary systems and 115 entries associated with visual triple systems or visual doubles where one component is an eclipsing, spectroscopic or astrometric binary. In Figure 9 we plot absolute V-band magnitude verses linear separation for the components of our remaining double/triple systems as compared to the outer HZ location. We note that in this plot, the binary components have brightnesses and separations that seem to avoid interference with the HZ, but this is an observational effect. Brighter objects are on average further away and therefore are not visually resolved at small separations, and spectroscopically detected binaries tend to "fill in" the space overlapping and interior to the HZ.

In order to assess the level of radiation variability within each habitable zone due to the constantly changing distance of a second star, we first calculated maximum and minimum distances of the second star to the habitable zone in question by again assuming a binary eccentricity of 0.67 and a semimajor axis as given by the observed linear separation. The fractional change in total HZ irradiance from maximum to minimum stellar separation was then calculated and systems where the HZ irradiance varies by more than 3% (the limit imposed in §2.2 for single variable stars) were eliminated. The largest binary separation that was cut during this part of the analysis was 86 AU. For triples, it was always the case that the third star was more than ~90 AU away from the close pair, so radiation variations were not a concern for this distant companion. Thus we eliminated 101 more Hipparcos entries from HabCat, and these systems are also shown in Figure 9.

One effect that we have not accounted for in this Section is that of terrestrial planet orbital eccentricity induced by the presence of a second star. For example, Mazeh, Krymolowski, & Roenfeld (1997) have suggested that the high eccentricity of the giant planet orbiting 16 Cyg B (e=0.63, a=1.6 AU) is caused by the presence of 16 Cyg A even given the wide separation of the pair (projected binary separation ~800 AU, mass ratio ~1, eccentricity similar to our assumption above, Cochran et al. 1997, Romanenko 1994). Recent work by Williams & Pollard (2002) suggests that Earths on fairly eccentric orbits might still be habitable, as long as the average stellar flux received over an entire orbit is comparable to that received by a planet on a circular orbit within the HZ. The expected planet eccentricity in binary systems (which will presumably be a function of the secondary mass, orbital eccentricity and inclination relative to the planet orbital plane) remains an outstanding question at this time.

§3.8.3 Spectroscopic and Astrometric Binaries

After all of the above cuts on binary stars, the remaining list contained 17,190 Hipparcos stars. From this list we identified 47 binaries with spectroscopically or astrometrically determined orbits. For each system a mass ratio of 0.23 (the average value according to Duquennoy & Mayor 1991) was assumed unless the system was a double-lined spectroscopic binary, in which case the mass ratios were calculated from the maximum velocity amplitudes of the two components. The eccentricities were directly available for spectroscopic binaries, and for astrometric binaries the eccentricity was assumed to be 0.67 as in the previous Section. For spectroscopic binaries the mass of the primary was estimated from the V-band magnitude of the system and the semimajor axis was then calculated from the period, while for astrometric binaries the projected semimajor axis was used. Outer HZ limits were calculated as above, and the inner HZ limits were estimated by subtracting an HZ width that is proportional to $L_*^{0.5}$ (where bolometric corrections were taken from Flower 1996) and equal to 0.81 AU for a 1 $L_\odot$ star. The entire HZ was required to be dynamically stable for a binary to be included in HabCat. Thus 12 SBs (all with periods less than 45 days) were found to have stable circumbinary habitable zones but none were found to have stable circumstellar HZs. Of the 29 astrometric binaries, only 9 were found to be dynamically habitable (with semimajor axes up to 0.5 AU), all of which support circumbinary habitable zones. Three stars had both spectroscopic and astrometric orbits, and they were required to be habitable according to all the observations. A total of 27 objects were removed in the Section.

§3.8.4 Extrasolar Giant Planets

Of the 17,163 stars remaining in HabCat, there are 55 stars known to host 65 total planets. With the smallest minimum planet mass of 0.12 Jupiter masses (HD 49674), all of these planets are likely to be gas giants and are thus unlikely to support Earth-like life. However, these planetary systems may still be habitable if (a) the giant planet does not interfere with the dynamic stability of the HZ, or (b) the giant planet occupies the HZ throughout its orbit, giving rise to potentially habitable moons. The potential habitability of moons is questionable, considering the high-radiation environment of a giant planet, possible gravitational focusing of large impactors by the giant planet, and the large eccentricities of most known extrasolar giant planets. Williams, Kasting & Wade (1997) have shown that the effects of radiation could be avoided if the moon has an Earth-like magnetic field, and work by Williams & Pollard (2002) suggests that planets in eccentric orbits will still be habitable, as long as the stellar flux averaged over one year at the moon's surface is comparable to that of a circular orbit (as was also mentioned in §3.8.2). The question of whether impact rates could remain high enough after a few billion years to prevent the appearance of complex life remains to be addressed.

In order to assess the dynamical habitability of these systems, we have located the inner and outer HZ limits as in the previous Section, and we have demanded that no giant planet come within three Hill radii of the habitable zone at any time during its orbit (hereafter referred to as the "Hill criterion"). At a separation of one Hill radius, the gravitational interaction between the giant planet and a terrestrial planet is approximately the same as that between either planet and the central star. This distance is calculated by:

$$R_H = a \, (M_p/3M_*)^{1/3}$$

where $M_p$ is the mass of the known giant planet, $M_*$ is the mass of the star, and $a$ is the giant planet's semimajor axis. Stellar masses were calculated by:

$$\log (M_*/M_\odot) = 0.48 - 0.105 \, M_{bol},$$

where $M_{bol}$ is the bolometric magnitude (bolometric corrections taken from Flower 1996). The data used in our calculations and habitability results (star name, mass, luminosity, planet mass, planet semimajor axis, inner HZ limit, outer HZ limit, and whether the system is habitable) are shown in Table 3.

In previous work, Jones et al. (2001) used the Hill criterion in addition to demanding that a terrestrial planet survive in the HZ for $10^8$ to $10^9$ years in numerical integrations for four exoplanetary systems, and Noble et al. (2002) have performed numerical integrations for three exoplanetary systems at higher time resolution but shorter total times. We have not performed any numerical integration, but we have used the Hill criterion to look at all of the 55 otherwise habitable exoplanetary systems. Figure 10 shows the location of habitable zones compared to the radial excursion of known planetary companions. Also indicated is the area falling within three times the giant planet's Hill radius. From this analysis we found that 17 of the known exoplanetary systems could host terrestrial planets in the habitable zone, 4 have giant planets orbiting completely within the HZ and could host habitable moons (HIPs 86796, 17096, 20723, and 8159), and we ruled out the other 34.

Where our list of exoplanetary systems overlaps with the objects analyzed by Jones et al. (2001, rho Crb, ups And, 47 UMa, Gliese 876) and Noble et al. (2002, HD 210277, 51 Peg, 47 UMa), we find that in every case our analysis reaches the same conclusions as the numerical integrations. We note that in this paper, for a system to be habitable the entire HZ must be stable, so we have not included 47 UMa in HabCat, although we agree with both authors that there are likely to be stable orbits in the inner HZ. We also find that the triple planetary system 55 Cnc is habitable, in agreement with work done by Marcy et al. (2002). After cutting these last 34 stars from our list, HabCat contains 17,129 habstars.

§4. THE CATALOG OF HABITABLE SYSTEMS

To briefly restate our criteria for habitability, a "habstar" must: (1) be at least 3 Gyr old, (2) be non-variable, (3) be capable of harboring terrestrial planets, and (4) support a dynamically stable habitable zone (defined by that annulus where an Earth-like planet could support liquid water on its surface). We have used those criteria to trim the list of 118,218 stars in the Hipparcos Catalogue down to the 17,129 stars in the Catalog of Nearby Habitable Stellar Systems (HabCat). HabCat will serve as the list of preferred targets for targeted searches carried out by the SETI Institute from the Allen Telescope Array.

Despite the broad array of data used to assemble this catalog, this exercise has forced us at every turn to admit that we are defining "habitability" from a position of considerable ignorance. A complete characterization of all the stars within a few hundred (or even a few tens of) parsecs, including their masses, ages, variability, and whether they have stellar companions or planetary systems (including terrestrial planets), is simply not realizable at this time. Additionally, many theoretical questions remain regarding the effects of metallicity on planet formation, the kinematics of stars and whether spiral arm crossings are truly deleterious to life forms, the effects of stellar variability (including timescales of hours, days and decades) on planet climate, the effect of stellar/giant planet companions on terrestrial planet orbital eccentricity, the effect of the stellar spectral energy distribution on the evolution of plants and other life forms, the suitability of giant planet moons for life (given expected impact rates, tidal heating, and particle radiation), etc. For SETI, this humbling situation is amplified when we consider that we have no indisputable definition for "life" itself, to say nothing of the precise conditions that are necessary and sufficient for life to evolve into a technological civilization detectable by a SETI search program. HabCat reflects the state of our current knowledge and will evolve as we learn more about Galactic structure, the Solar Neighborhood, planets, life in the Solar System, and the evolution of intelligence on Earth.

Figure 11 shows habstar distances as a function of spectral type as well as the cumulative distribution for all spectral types. Distances have been corrected for the Lutz-Kelker bias using the method outlined in Hanson (1979) and the parameters specific to the Hipparcos sample as determined by Reid (1997). The Allen Telescope Array (ATA) will have access to the 12,319 habstars north of −34° in declination. With predicted ATA sensitivities, we will be able to place upper limits of $1.2 \times 10^{13}$ Watts EIRP (comparable to the Arecibo planetary radar power) on ETI transmissions at a distance of 300pc. As can be seen in Figure 12, the Hipparcos limiting magnitude and our cut on early-type stars have conspired to create a list comprised mostly of late F and G stars (44% F, 39% G, 14% K, and 3% M stars). The list of Hipparcos stars contained in HabCat is available in the electronic version of this paper and via email from M. Turnbull, and a short sample version of the catalog is shown in Table 4.

## ACKNOWLEDGEMENTS

We are deeply grateful for many helpful discussions and data provided by Brian Skiff, Todd Henry, Geoff Marcy, Mark Giampapa, Alan Boss, Neill Reid, Darren Williams, Jim Kasting, Laurance Doyle, Eric Mamajek, Patrick Young, Betsy Green, Don McCarthy, Jonathan Lunine, Jim Liebert, Patrick Guillout, Dimitri Pourbaix, Guillermo Gonzalez, Dennis Zaritsky and Bill Tifft, and for computer support from Tim Pickering, Chris Groppi, Tucker Bradford, Ken Okutake, and Steve Brockbank. We would also like to express gratitude to all those who maintain the online databases (ADS, SIMBAD, VIZIER, etc.), without whom the making of HabCat would not have been possible. This work was funded by the NASA Arizona Space Grant Graduate Fellowship, the NSF Collaborative Advancement for Teaching Technology and Science Graduate Fellowship, the NSF Integrated Graduate Education, Research and Training (IGERT) Fellowship, and the SETI Institute. MT is also grateful to the UC-Berkeley Astronomy Department for its hospitality during Summer 2002.

Table 1
The Celestia Query

| Query Parameter | Specification | Number of Stars |
|---|---|---|
| 1. Hipparcos Stars | All entries | 118,218 |
| 2. Photometry | $-1.037 < B-V < 5.460$ | 116,937 |
| 3. Parallax | $\pi > 0$ mas | 113,710 |
| 4. Parallax Uncertainty | $\sigma_\pi/\pi < 0.3$ | 69,301 |
| 5. Coarse Variability | < 0.06 mag | 4112 |
| 6. Coarse Variability | 0.06 to 0.6 mag | 6351 |
| 7. Coarse Variability | > 0.6 mag | 1099 |
| 8. Variability Annex | Unsolved variables (2) | 5542 |
| 9. Variability Annex | Light curve (not folded) (C) | 827 |
| 10. Variability Type (1 Letter) | Microvariable (M) | 1045 |
| 11. Variability Type (1 Letter) | Unsolved variables (U) | 7784 |
| 12. Multiplicity Annex | Variability induced movers (V) | 288 |
| 13. Multiplicity Annex | Stochastic solution (X) | 1561 |
| 14. Resolved Components | 3 or 4 | 135 |
| 15. Variability Type (5 Letters) | E, EA, EB, EW | 986 |
| 16. Combined Criteria | 1 *AND* 2 *AND* 3 *AND* 4 | 69,014 |
| 17. Combined Criteria | *NOT* 14 *AND* (4 OR 5 OR 6) | 10,576 |
| 18. Combined Criteria | 1 *AND NOT* (7 OR 8 OR 9 OR 10 OR 11 OR 12 OR 13 OR 16) | 64,120 |

Table 2
Hipparcos Binary/Multiple Stars

| Multiplicity Type | EB | CCDM | SB | SNS | O | G |
|---|---|---|---|---|---|---|
| EB...................... | 19 | ... | ... | ... | ... | ... |
| CCDM................ | 3 | 2433 | ... | ... | ... | ... |
| SB...................... | 4 | 32 | 49 | ... | ... | ... |
| SNS................... | 1 | 122 | 2 | 627 | ... | ... |
| O........................ | 0 | 7 | 6 | 0 | 37 | ... |
| G........................ | 0 | 25 | 3 | 0 | 0 | 542 |

Table 3
Extrasolar Giant Planet Data

| Star parameters | | | | | Planet parameters | | | HZ parameters | | | |
|---|---|---|---|---|---|---|---|---|---|---|---|
| HIP | Other | $M_v$ | $M_*/M_\odot$ | $L_*/L_\odot$ | $M \sin i$ $M_J$ | $a$ AU | $e$ | $a_{out}$ AU | $a_{in}$ AU | HAB1[1] | HAB2[2] |
| 522  | HD142   | 3.66 | 1.25 | 2.77 | 1.36 | 0.98 | 0.37 | 2.46 | 1.09 | NO     | NO     |
| 1931 | HD2039  | 4.23 | 1.11 | 1.75 | 5.10 | 2.20 | 0.69 | 2.07 | 0.99 | NO     | NO     |
| 3502 | HD4203  | 4.24 | 1.13 | 1.86 | 1.64 | 1.09 | 0.53 | 2.07 | 0.95 | NO     | NO     |
| 6643 | HD8574  | 3.90 | 1.19 | 2.28 | 2.23 | 0.76 | 0.40 | 2.29 | 1.05 | NO     | NO     |
| 7513 | Ups And | 3.45 | 1.31 | 3.36 | 0.68 | 0.06 | 0.01 | 2.60 | 1.10 | YES    | NO     |
| 7513 | Ups And | 3.45 | 1.31 | 3.36 | 1.94 | 0.83 | 0.25 | 2.60 | 1.10 | NO     | NO     |
| 7513 | Ups And | 3.45 | 1.31 | 3.36 | 4.02 | 2.54 | 0.31 | 2.60 | 1.10 | NO     | NO     |
| 8159 | HD10697 | 3.71 | 1.27 | 2.94 | 6.08 | 2.12 | 0.11 | 2.42 | 1.01 | YES[3] | YES[3] |

[1] "YES" indicates that the HZ is stable given the orbit of the planet in question.
[2] "YES" indicates that the HZ is stable given the orbits of all planets in the system.
[3] Moons in this system may be habitable

| Star parameters | | | | | Planet parameters | | | HZ parameters | | | |
|---|---|---|---|---|---|---|---|---|---|---|---|
| HIP | Other | $M_v$ | $M_*/M_\odot$ | $L_*/L_\odot$ | M sin i $M_J$ | a AU | e | $a_{out}$ AU | $a_{in}$ AU | HAB1[1] | HAB2[2] |
| 9683 | HD12661b | 4.58 | 1.04 | 1.31 | 2.30 | 0.82 | 0.35 | 1.87 | 0.93 | NO | NO |
| 9683 | HD12661c | 4.58 | 1.04 | 1.31 | 1.56 | 2.56 | 0.20 | 1.87 | 0.93 | NO | NO |
| 12048 | HD16141 | 4.05 | 1.16 | 2.07 | 0.22 | 0.35 | 0.00 | 2.18 | 1.00 | YES | YES |
| 14954 | HD19994 | 3.32 | 1.36 | 3.87 | 1.66 | 1.19 | 0.20 | 2.71 | 1.09 | NO | NO |
| 17096 | HD23079 | 4.42 | 1.05 | 1.40 | 2.76 | 1.48 | 0.14 | 1.96 | 0.99 | YES[3] | YES[3] |
| 20723 | HD28185 | 4.81 | 0.99 | 1.08 | 5.70 | 1.03 | 0.07 | 1.75 | 0.89 | YES[3] | YES[3] |
| 21850 | HD30177 | 4.72 | 1.01 | 1.20 | 7.64 | 2.65 | 0.21 | 1.80 | 0.90 | NO | NO |
| 24205 | HD33636 | 4.71 | 0.99 | 1.08 | 7.71 | 2.62 | 0.39 | 1.80 | 0.95 | NO | NO |
| 31246 | HD46375 | 5.29 | 0.90 | 0.76 | 0.25 | 0.04 | 0.02 | 1.52 | 0.81 | YES | YES |
| 32916 | HD49674 | 5.05 | 0.93 | 0.86 | 0.12 | 0.06 | 0.00 | 1.63 | 0.87 | YES | YES |
| 33212 | HD50554 | 4.38 | 1.06 | 1.46 | 3.72 | 2.32 | 0.51 | 1.99 | 0.99 | NO | NO |
| 33719 | HD52265 | 4.05 | 1.15 | 1.97 | 1.14 | 0.49 | 0.29 | 2.19 | 1.03 | YES | YES |
| 40687 | HD68988 | 4.35 | 1.08 | 1.55 | 1.90 | 0.07 | 0.14 | 2.00 | 0.98 | YES | YES |
| 42030 | HD72659 | 3.91 | 1.19 | 2.29 | 2.54 | 3.24 | 0.18 | 2.28 | 1.04 | NO | NO |
| 42723 | HD74156b | 3.56 | 1.29 | 3.12 | 1.55 | 0.28 | 0.65 | 2.52 | 1.07 | YES | NO |
| 42723 | HD74156c | 3.56 | 1.29 | 3.12 | 7.46 | 3.47 | 0.40 | 2.52 | 1.07 | NO | NO |
| 43177 | HD75289 | 4.04 | 1.15 | 1.99 | 0.46 | 0.05 | 0.01 | 2.19 | 1.03 | YES | YES |
| 43587 | 55 Cnc | 5.47 | 0.87 | 0.65 | 0.84 | 0.12 | 0.02 | 1.45 | 0.79 | YES | YES |
| 43587 | 55 Cnc | 5.47 | 0.87 | 0.65 | 0.21 | 0.24 | 0.34 | 1.45 | 0.79 | YES | YES |
| 43587 | 55 Cnc | 5.47 | 0.87 | 0.65 | 4.05 | 5.90 | 0.16 | 1.45 | 0.79 | YES | YES |
| 43686 | HD76700 | 4.28 | 1.12 | 1.77 | 0.19 | 0.05 | 0.00 | 2.05 | 0.95 | YES | YES |
| 47007 | HD82943c | 4.35 | 1.08 | 1.54 | 0.88 | 0.73 | 0.54 | 2.01 | 0.99 | NO | NO |
| 47007 | HD82943b | 4.35 | 1.08 | 1.54 | 1.63 | 1.16 | 0.41 | 2.01 | 0.99 | NO | NO |
| 50786 | HD89744 | 2.78 | 1.53 | 6.25 | 7.17 | 0.88 | 0.70 | 3.17 | 1.12 | NO | NO |
| 52409 | HD92788 | 4.76 | 0.99 | 1.09 | 3.88 | 0.97 | 0.28 | 1.78 | 0.92 | NO | NO |

| Star parameters | | | | | Planet parameters | | | HZ parameters | | | |
|---|---|---|---|---|---|---|---|---|---|---|---|
| HIP | Other | $M_v$ | $M_*/M_\odot$ | $L_*/L_\odot$ | M sin i $M_J$ | a AU | e | $a_{out}$ AU | $a_{in}$ AU | HAB1[1] | HAB2[2] |
| 53721 | 47 Uma | 4.29 | 1.09 | 1.62 | 2.56 | 2.09 | 0.06 | 2.04 | 0.99 | NO | NO |
| 53721 | 47 Uma | 4.29 | 1.09 | 1.62 | 0.76 | 3.78 | 0.00 | 2.04 | 0.99 | YES | NO |
| 59610 | HD106252 | 4.54 | 1.03 | 1.30 | 6.79 | 2.53 | 0.57 | 1.89 | 0.96 | NO | NO |
| 60644 | HD108147 | 4.06 | 1.14 | 1.92 | 0.41 | 0.08 | 0.20 | 2.18 | 1.04 | YES | YES |
| 61028 | HD108874 | 4.58 | 1.04 | 1.33 | 1.65 | 1.07 | 0.20 | 1.87 | 0.92 | NO | NO |
| 64457 | HD114783 | 6.01 | 0.78 | 0.41 | 0.99 | 1.20 | 0.10 | 1.24 | 0.71 | NO | NO |
| 65721 | 70 Vir | 3.68 | 1.27 | 3.00 | 7.41 | 0.48 | 0.40 | 2.44 | 1.01 | YES | YES |
| 72339 | HD130322 | 5.67 | 0.82 | 0.50 | 1.15 | 0.09 | 0.05 | 1.37 | 0.78 | YES | YES |
| 74500 | HD134987 | 4.42 | 1.07 | 1.49 | 1.63 | 0.82 | 0.37 | 1.96 | 0.96 | NO | NO |
| 74948 | HD136118 | 3.34 | 1.35 | 3.75 | 11.91 | 2.39 | 0.37 | 2.69 | 1.10 | NO | NO |
| 77740 | HD141937 | 4.63 | 1.01 | 1.19 | 9.67 | 1.48 | 0.40 | 1.85 | 0.95 | NO | NO |
| 78459 | Rho Crb | 4.18 | 1.12 | 1.78 | 0.99 | 0.22 | 0.07 | 2.10 | 1.01 | YES | YES |
| 79248 | HD145675 | 5.32 | 0.90 | 0.75 | 3.90 | 2.87 | 0.37 | 1.51 | 0.80 | NO | NO |
| 86796 | HD160691 | 4.20 | 1.13 | 1.83 | 1.74 | 1.48 | 0.31 | 2.09 | 0.98 | YES[3] | YES[3] |
| 89844 | HD168443b | 4.03 | 1.18 | 2.19 | 7.64 | 0.30 | 0.53 | 2.20 | 0.98 | YES | NO |
| 89844 | HD168443c | 4.03 | 1.18 | 2.19 | 16.96 | 2.87 | 0.20 | 2.20 | 0.98 | NO | NO |
| 90004 | HD168746 | 4.78 | 0.99 | 1.09 | 0.24 | 0.07 | 0.00 | 1.77 | 0.91 | YES | YES |
| 90485 | HD169830 | 3.10 | 1.42 | 4.62 | 2.95 | 0.82 | 0.34 | 2.89 | 1.12 | NO | NO |
| 96901 | 16 Cyg B | 4.60 | 1.02 | 1.25 | 1.68 | 1.69 | 0.68 | 1.86 | 0.95 | NO | NO |
| 97336 | HD187123 | 4.43 | 1.06 | 1.46 | 0.54 | 0.04 | 0.01 | 1.96 | 0.97 | YES | YES |
| 98714 | HD190228 | 3.33 | 1.40 | 4.36 | 5.01 | 2.25 | 0.43 | 2.70 | 0.98 | NO | NO |
| 101806 | HD196050 | 4.14 | 1.14 | 1.91 | 2.81 | 2.41 | 0.20 | 2.13 | 0.99 | NO | NO |
| 104903 | HD202206 | 4.75 | 1.00 | 1.12 | 14.68 | 0.77 | 0.42 | 1.78 | 0.91 | NO | NO |
| 108859 | HD209458 | 4.29 | 1.09 | 1.60 | 0.63 | 0.05 | 0.02 | 2.04 | 1.00 | YES | YES |
| 109378 | HD210277 | 4.90 | 0.97 | 1.02 | 1.29 | 1.12 | 0.45 | 1.71 | 0.88 | NO | NO |
| 111143 | HD213240 | 3.76 | 1.23 | 2.61 | 4.49 | 2.02 | 0.45 | 2.38 | 1.05 | NO | NO |

| Star parameters | | | | | Planet parameters | | | HZ parameters | | | |
|---|---|---|---|---|---|---|---|---|---|---|---|
| HIP | Other | $M_v$ | $M_*/M_\odot$ | $L_*/L_\odot$ | M sin i $M_J$ | a AU | e | $a_{out}$ AU | $a_{in}$ AU | HAB1[1] | HAB2[2] |
| 113020 | GJ876c | 11.80 | 0.27 | 0.01 | 0.56 | 0.13 | 0.27 | 0.27 | 0.20 | YES | NO |
| 113020 | GJ876b | 11.80 | 0.27 | 0.01 | 1.89 | 0.21 | 0.10 | 0.27 | 0.20 | NO | NO |
| 113137 | HD216437 | 3.92 | 1.20 | 2.32 | 2.09 | 2.38 | 0.34 | 2.27 | 1.02 | NO | NO |
| 113357 | 51 Peg | 4.52 | 1.04 | 1.35 | 0.46 | 0.05 | 0.01 | 1.91 | 0.95 | YES | YES |
| 113421 | HD217107 | 4.70 | 1.01 | 1.20 | 1.29 | 0.07 | 0.14 | 1.81 | 0.91 | YES | YES |
| 116906 | HD222582 | 4.57 | 1.03 | 1.28 | 5.20 | 1.36 | 0.76 | 1.88 | 0.95 | NO | NO |

figure 1. The color-magnitude diagram for the Celestia sample, showing evolutionary tracks up to 3 Gyr for a 1.4 Solar mass star (yellow) and a 1.6 Solar mass star (cyan) with Solar metallicity. Also shown are the cuts applied below the main sequence and at the Hertzsprung gap, and a maximum luminosity cut at $M_v = 2.5$ (described in §3.1). The shaded area of the color-magnitude diagram was excluded from HabCat.

Figure 2. The distribution of calcium H&K activity for 1408 stars in our sample (left axis). The vertical dashed line indicates our cutoff for "young" stars at log $R'_{HK} = -4.75$, and the dotted line indicates the CE-age relation from Donahue (1993) in gigayears (right axis) as a function of log $R'_{HK}$. The Sun's excursion in log $R'_{HK}$ from Maunder Minimum (−5.1) to Solar Maximum (−4.75) is also shown.

Figure 3. The number of RASS detections as a function of distance (left axis). The dotted line indicates the maximum age star detectable to *ROSAT* (right axis) at the given distance, based on the $L_x$-age relation from Güdel et al. (1997). The vertical dashed line indicates the distance within which detected stars may be older than 3 Gyr.

Figure 4. (a) U, V and W velocity components (circles, squares, and triangles, respectively) as a function of metallicity for the 189 stars in Edvardsson et al. (1993). The vertical dashed line at [Fe/H] = −0.4 denotes the minimum metallicity cutoff applied to HabCat.

Figure 5. (a) The V and U components of galactic space velocity for 20,765 potential habstars resulting from the cuts of §3.6. Stars with 3-dimensional space velocities that indicate "low" metallicity are black, and kinematically "high" metallicity stars are indicated in gray. (b) The same, for V and W components of galactic space velocity. V is along the direction of galactic rotation, U is positive toward the galactic anti-center, and W is perpendicular to the galactic plane.

Figure 6. The metallicity distributions of kinematically low metallicity stars (KLM, filled histogram) and kinematically high metallicity stars (KHM, open histogram). The metallicities were derived either from spectroscopic data or photometry (described in §3.7.3), and the vertical dashed line denotes the minimum metallicity cutoff applied to HabCat.

Figure 7. Spectroscopically determined metallicities ($[Fe/H]_{spec}$, from Cayrel de Strobel et al. 2001) compared to photometrically determined metallicities using Strömgren uvby photometry ($[Fe/H]_{phot}$, see Olsen et al. references) for 602 Hipparcos stars. $[Fe/H]_{phot}$ agrees well enough with $[Fe/H]_{spec}$ for use in coarsely assigning stars to "low" ($[Fe/H] < -0.4$) and "high" ($[Fe/H] \geq -0.4$) metallicity categories. The dashed lines denote the minimum metallicity cutoff applied to HabCat.

Figure 8. The location of possible habitable zones (shaded area) and stable planet orbits for (a) a binary system and (b) a triple system (not to scale). Stable S-type (circumstellar) orbits exist interior to the dotted lines, and stable P-type (circumbinary) orbits exist exterior to the solid lines. Note that while the Figure illustrates both circumbinary and circumstellar HZs, in reality a circumbinary HZ can only exist when two stars are so close together that neither could support a circumstellar HZ.

Figure 9. The location of the outer HZ limit as a function of absolute V magnitude (dashed line), compared to the separations of the binaries (squares) and triple systems (stars) of § 3.8.2. Black symbols denote systems that were cut due to dynamical instability of the HZ, magenta symbols denote systems that were cut due to radiative variations in the HZ of greater than 3%, and cyan symbols denote systems that were kept in HabCat.

Figure 10. The location of the HZ (shown in magenta) as a function of absolute V magnitude for stars known to harbor extrasolar giant planets, compared to the total orbital excursions of the exoplanets (solid black line) and the orbital excursion plus three times the Hill sphere (cyan dashed line).

Figure 11. The number of HabCat stars as a function of distance for M stars (solid histogram), K stars (dark-hatched histogram), G stars (light-hatched histogram), F stars (horizontal-lined histogram), and all stars (open histogram). The furthest star in HabCat is at ~300 pc, and transmitter powers comparable to the Arecibo planetary radar will be detectable to the ATA at this distance.

Figure 12. The number of HabCat stars as a function of B−V color. Our spectral type cutoff at ~F5 and the Hipparcos limiting magnitude result in a sample that contains primarily late F- and early G-type stars.